%% file: atomicflashmem.tex
\documentclass{intech}

\usepackage{graphicx}% Include figure files
\usepackage{amsmath}
\newcommand{\gst}{Ge$_{2}$Sb$_{2}$Te$_{5}$}
\newcommand{\Agi}{(GeSe$_{3}$)$_{0.9}$Ag$_{0.1}$}
\newcommand{\Agii}{(GeSe$_{3}$)$_{0.8}$Ag$_{0.2}$}
\newcommand{\Agiii}{(GeSe$_{3}$)$_{0.85}$Ag$_{0.15}$}
\newcommand{\AgCu}{(GeSe$_{3}$)$_{0.8}$Cu$_{0.1}$Ag$_{0.1}$}
\newcommand{\Cui}{(GeSe$_{3}$)$_{0.9}$Cu$_{0.1}$}
\newcommand{\Cuiii}{(GeSe$_{3}$)$_{0.8}$Cu$_{0.2}$}
\newcommand{\Cuii}{(GeSe$_{3}$)$_{0.77}$Cu$_{0.03}$Ag$_{0.2}$}

\booktitle{Will-be-set-by-IN-TECH}%

\chaptertitle{Atomistic Simulations of Flash Memory Materials Based on Chalcogenide Glasses } % You know your chapter title?

\authors{Bin Cai, Binay Prasai and D. A. Drabold}
\affiliation{Department of Physics and Astronomy, Ohio University, Athens, Ohio 45701}
\country{United States}

\begin{document}

\maketitle

\section{Introduction}

In the last decade, the market for non-volatile computer storage has experienced rapid growth. However, as the device size scales down, currently used NOR and NAND technology is facing limitations from tunnel oxide and electrostatic interactions between cells \citep{Lac08pps}. To solve the scaling problems, in the last fifteen years, a number of alternative Flash memory technologies have been proposed and studied. The phase-change memory, solid electrolyte memory, ferroelectric memory, magnetic memory and molecular memory using conducting molecules and carbon nanotubes all are promising candidates to replace technologies that are reaching their limit \citep{Dad09EPJB,ChungNanotch10}.

Vitreous materials involving chemical species from column VI other than oxygen are called Chalcogenide glasses. Chalcogenide glasses have long held interest for applications such as infrared detectors and optical fibers. Recently, chalcogenide glasses attracted further attention due to their promising application in data storage devices. Two of the examples are phase-change memory materials based on tellurium alloys and solid electrolyte memory materials based on metal-doped Ge-Se glasses. In this chapter, by using \textit{ab-initio} molecular dynamics, we introduce the latest simulation results on two materials for flash memory devices: \gst ~and Ge-Se-Cu-Ag. This chapter is a review of our previous work including some of our published figures and text in ~\cite{caigst10} and ~\cite{prasai} and also includes several new results. We organized the chapter as follows: we first introduce the simulation method used in all calculations in section~\ref{simu}; in section~\ref{GST}, we show the key findings on \gst\citep{caigst10}; then we switch to the study on Ge-Se-Cu-Ag in section~\ref{gese} \citep{prasai}. For both cases, after forming realistic atomistic models, we analyze the topology and electronic structure, predict their properties, and compare with experimental results.

\section{Simulation Method} \label{simu}

We use the \textit{ab-initio} molecular dynamics method (ab-MD) to generate atomistic models. When such schemes are applied, the initial position of atoms are usually randomized and the system is then annealed or relaxed to seek suitable local energy minima. The final model will be the one with a minimum total energy and agreeing with experimental measurements. One of the popular techniques based on ab-MD is the "Cook and Quench" method, for which the MD simulation is performed at a temperature well above melting point, which will force the system to lose memory of the initial configuration. Finally, the system is equilibrated at a lower temperature, like room temperature. Then an energy minimization is applied. Many realistic models are made by such a simple but powerful method.

In our work, all of the calculations were carried out using periodic boundary conditions with the Vienna Ab-initio Simulation Package(VASP). VASP is based on density functional theory using a plane wave basis \citep{vasp}. For \gst, we used the projector augmented-wave (PAW) potentials and generalized gradient approximation PBE (GGA-PBE) method \citep{vasppaw,vasppbe}; for Ge-Se-Cu-Ag, we used the local density approximation (LDA) for the exchange correlation energy in conjunction with the Vanderbilt Ultra Soft pseudopotentials \citep{vasplda,vaspuu}. Both systems were annealed, equilibrated and cooled using molecular dynamic (MD) option of VASP and relaxation is carried out in conjugate gradient (CG) mode. Moreover, to obtain a better estimation for electronic gap, we applied Hartree-Fock (HF) calculation when analyzing the electronic structure of amorphous \gst. Though HF is known to exaggerate both the optical gap and charge fluctuation in the electron gas. These features are helpful to us for diagnosing the correlations between topology and electronic properties.

\section{Phase-Change Memory Material} \label{GST}
\input{APLGST}

\input{electrolyte}

\section{Conclusion}

By using molecular dynamic simulations, we generate the atomistic models of \gst ~and Ge-Se-Ag-Cu and analyzed their topology and electronic structures. Both phase-change and electrolyte solid materials show the promising properties as candidates to replace the contemporary technologies in Flash memory. With further development, we believe the new generation of the computer storage device will eventually appear with much smaller size, higher speed and more reliable features. We show that computer simulation can lend insight into promising technologies.

\section{Acknowledgement}
The authors would like to particularly thank Professor S.R. Elliott and Dr. J. Hegedus for an introduction to phase-change memory materials and collaboration on GST work. DAD thanks Professor M. Mitkova and M. Kozicki for years of advice and collaboration. The authors also want to thank Prof. Gang Chen and Dr. Mingliang Zhang for their assistance and suggestions. This work was partly supported by the US NSF grant DMR-09033225, DMR-0844014 and DMR-0903225.

\newpage
%%% Without a bib file, write your references like this: **************

%%% With a bib file, include it! *************************************
%\bibliography{temp}

\end{document}

%% file: APLGST.tex
%\usepackage{graphicx}% Include figure files
%\usepackage{dcolumn}% Align table columns on decimal point
%\usepackage{bm}% bold math
%\usepackage{amssymb}
%\newcommand{\gst}{Ge$_{2}$Sb$_{2}$Te$_{5}$}
%\usepackage[mathlines]{lineno}% Enable numbering of text and display math
%\linenumbers\relax % Commence numbering lines

\subsection{Background}

For Ge-Sb-Te (GST) alloys, there exists a rapid and reversible transition between crystalline and amorphous states. Controlled modification of electrical conductivity and optical properties of the transition is the basis for promising FLASH and optical memory devices. Akola and Jones \citep{Akola07prb} analyzed the structure of liquid and amorphous phases, and compared the electronic structure with the crystal phase. In 2008, Hegedus and Elliott \citep{Hed08} reproduced the crystal-amorphous transition by MD simulation, and they found that the rapid crystal growth was due to the presence of crystal fragments -- four member square rings (so-called "seeds") in amorphous and liquid phases. Their work provided a way to track the dynamic changes of network topology and electronic structure at the same time. Welnic and co-workers \citep{Welnic07prl} studied the origin of optical properties and argued that the optical contrast between amorphous and crystalline phases is due to a change in local order of Ge atoms. Despite this progress, the correlation between topology and electronic structure, most especially the origin of the change in the electronic gap, is still imperfectly understood. One of the challenges is the basic limitation of the LDA for estimating the gap.

\subsection{Model Preparation}

We began our work by creating amorphous \gst ~models by using the Vienna \textit{Ab-initio} Simulation Package (VASP)-- a plane-wave DFT code, using a PAW potential and the GGA-PBE method \citep{Hed08}. 63-atom amorphous \gst ~models with lattice constant 12.5 \AA ~were made as follows. The system was first melted and equilibrated at 1000K, followed by a rapid quench to 500K with a quench rate of 16K/ps. Then the system was equilibrated for 20 ps and data collection began at 10ps. For the crystal phase, 108-atom crystal \gst ~cells with lattice constant 21.316\AA ~are generated based on NaCl rock-salt structure with 10\% vacancies: 60 Te atoms occupied the Na sites; 24 Ge atoms and 24 Sb atoms randomly occupied the Cl sites which left 10 Cl sites unoccupied. The system was then relaxed under zero-pressure till the minimum total energy was obtained.

\subsection{Result and Discussions}

\subsubsection{Bond Statistics}
\begin{table}[tbp]
%\begin{center}
\centering
\caption {Mean coordinations, bond types and seeds(four member square rings) statistics at 500K. The result obtained at 1000K is listed in brackets (coordination cutoff=3.2{\AA}).}
\label{table1}
%\begin{ruledtabular}
\begin{tabular}{ccccccccc}
&$N_{tot}$&$N_{Te}$&$N_{Ge}$&$N_{Sb}$& \\
\hline
Te&3.4(3.0)&20\%(30\%)&47\%(41\%)&33\%(29\%)\\
Ge&4.6(4.3)&86\%(71\%)&5\%(13\%)&9\%(16\%)\\
Sb&4.1(3.6)&69\%(62\%)&11\%(20\%)&20\%(18\%)\\
$N_{seed}$&18(1.8)&52\%(10\%)&69\%(12\%)&53\%(10\%)\\
\end{tabular}
%\end{ruledtabular}
%\end{center}
\end{table}

\begin{figure}[htbp]
%\begin{center}
\centering
\resizebox{1.0\columnwidth}{!}{
\includegraphics{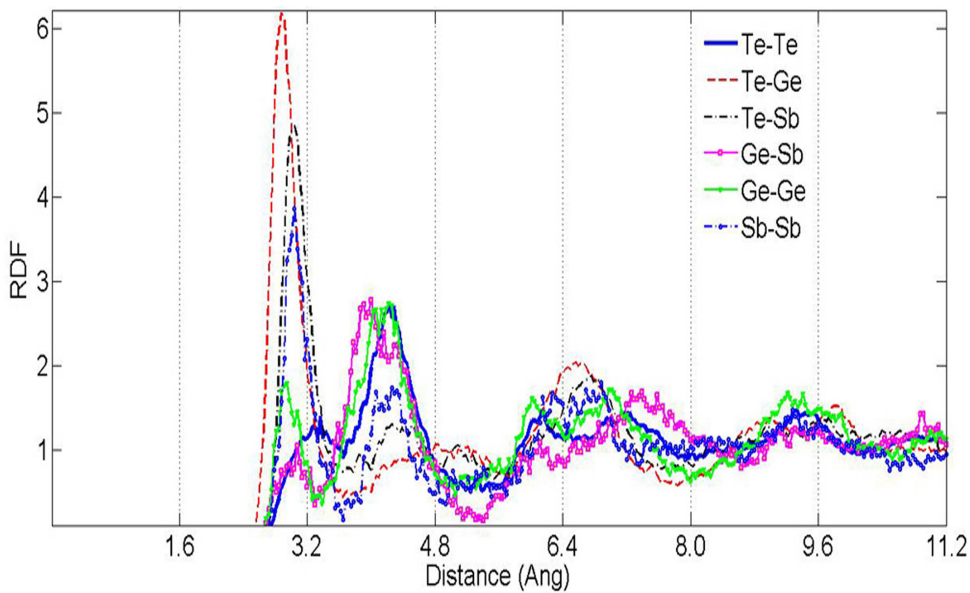}}
\caption{\label{figrdf}Partial radial distribution functions for a-\gst.}
%\end{center}
\end{figure}

The calculated atomic coordinations for a-\gst (500K) and l-\gst (1000K) are listed in Table 1 with a 3.2{\AA} cut-off. These results are similar to \cite{Akola07prb}, although in our case, the mean coordination of Ge atoms is slightly increased after thermal quench and equilibration, which may be due to the higher equilibration temperature used and/or size artifacts for our smaller model. More highly-coordinated Ge and Sb (5-fold,6-fold) atoms appeared in the amorphous phase, which suggests that a near-octahedral structure may be formed (square-rings and 8-atom cubes). These results indicate that structural ordering is enhanced in the amorphous phase relative to the liquid. Moreover, the number of wrong bonds (Te-Te,Ge-Ge,Sb-Sb and Ge-Sb) are decreased from 1000K to 500K which indicates that the chemical order is also improved. The average number of "seeds" (four member square rings) shows an increase in the amorphous phase and more than 50\% of the atoms are involved in "seeds", compared to only 10\% in the liquid phase. The calculated partial radial distribution functions are plotted in Fig.~\ref{figrdf}. The first peak in the Te-Ge and Te-Sb partials are located at 2.81{\AA} and 2.92{\AA}. The shallow first minima imply that the coordination is sensitive to the cutoff value selected. The Te-Ge partial has a broad and weak second peak. However, the Te-Sb partial possesses a second peak with a maximum at 4.4{\AA} which indicates that Ge and Sb atoms differ in local environment relative to Te atoms. Regarding the homopolar bonds, there is a major peak for the Sb-Sb partial centered at 2.9{\AA}. These results are similar to other simulations \citep{Akola07prb} and also experimental results \citep{Natio10}.

\subsubsection{Electronic Structure}

\begin{figure}[htbp]
%\begin{center}
\centering
\resizebox{1.0\columnwidth}{!}{
\includegraphics{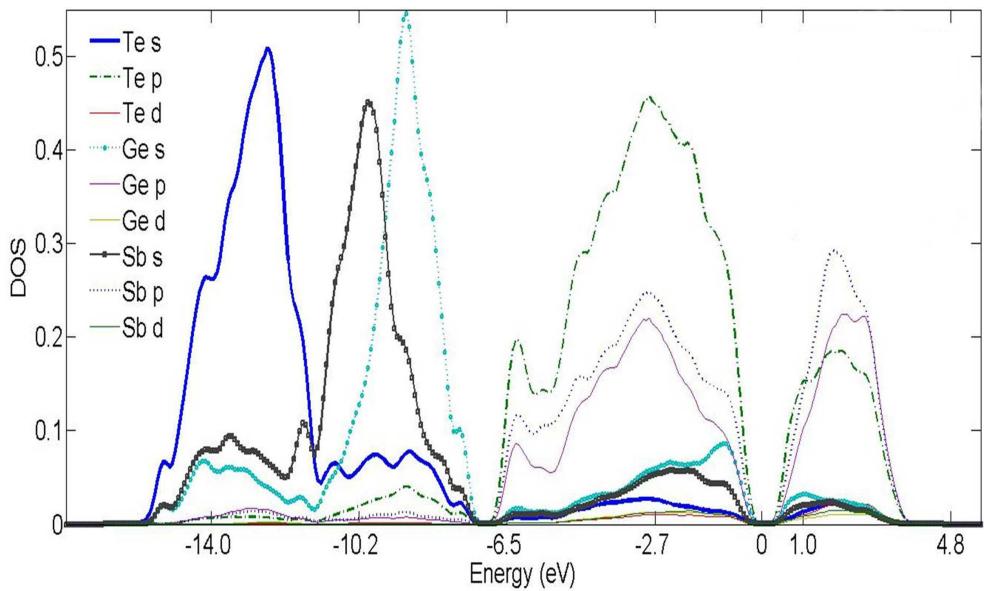}}
\caption{\label{figEtype}Electronic densities of states projected onto different atomic species and orbitals. The Fermi level is at 0 eV.}
%\end{center}
\end{figure}

The electronic structure is analyzed through the electronic density of states (EDOS) obtained from Hartree-Fock (HF) calculations. HF is used only to analyze the EDOS, not for forces and total energies. HF is known to exaggerate both the optical gap and charge fluctuations in the electron gas. These features are helpful to us here for diagnosing structural correlations. In the following discussion, the calculated EDOS is averaged over 1000 configurations from the last 2 ps when the cell is in thermal equilibrium at a fixed temperature of 500K. Finally, the averaged HF result of the amorphous phase gives an electronic gap around 0.4eV which is wider than the DFT result--0.2eV \citep{Akola07prb} and is closer to the experimental value--0.7eV \citep{Lee05expgap}. Although the gap is still smaller than the experimental value, it is much improved over LDA, and this may imply that HF provides a better starting point for analysis of the electronic structure.

\begin{figure}[htbp]
%\begin{center}
\centering
\resizebox{0.9\columnwidth}{!}{
\includegraphics{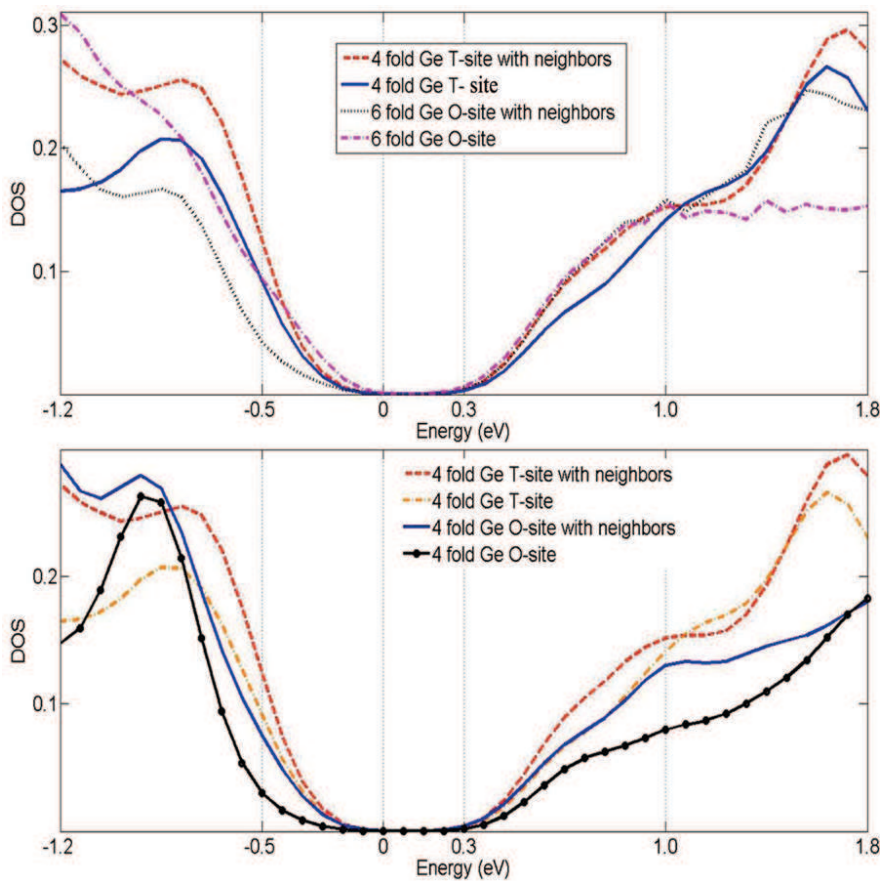}}
\caption{\label{figTO}Projected EDOS on Ge atoms at tetrahedral and octahedral sites. ``T" and ``O" represent tetrahedral and octahedral sites. The ``Ge-T/O site" plot only considers the contribution of Ge atoms to the EDOS, while the ``Ge-T/O site with neighbors" plot contains the contribution of Ge atoms and their neighbors. The Fermi level is at 0 eV.}
%\end{center}
\end{figure}

To correlate topology with electronic structure, we projected the EDOS onto different local sites and are able to attribute the electronic states to specific structural units. We first show the averaged EDOS for different species and orbitals in Fig.~\ref{figEtype}. The key findings are that, for all three species, p orbitals dominate the gap and tail states; if considering the species, Te-p,Sb-p,Ge-p,Ge-s and Sb-s are important in determining tail states and the magnitude of the gap (Fig.~\ref{figEtype}). To further correlate structural oddities with electronic states, we also sort atoms with specific features into different groups and accumulate the contribution to EDOS. We briefly report that groups forming homopolar or heteropolar bonds showed that there is a significant difference at a ``deep gap" around -7eV below the Fermi level (0eV) in EDOS (atoms involved in heteropolar bonds form a bigger deep gap); however, atoms forming homopolar bonds have a minor impact on tail states and the electronic gap near the Fermi level. Considering the coordination, for Te, 2-fold Te atoms contribute to a narrowed gap and conduction-band tail states appear; for Ge atoms, the contributions for 3, 4, 5 and 6-fold atoms are almost the same; for Sb atoms, the conduction-band tail of 6-fold Sb atoms is pushed to a low-energy level and the valence-band tail associated with 3-fold Sb atoms which satisfy the ``8-N" rule is pushed into a higher energy region. While there are differences in electronic tail states and the gap value associated with coordination numbers, the influence is fairly weak. Similarly, sorting atoms involved in ``seeds" or not also showed a minor impact on gap magnitude and tail states.

We also considered the ``umbrella flip" of Ge atoms. We compared the EDOS of Ge atoms sitting at octahedral sites (O-site) and tetrahedral sites (T-site), as we illustrate in Fig.~\ref{figTO}. The projected EDOS on Ge atoms and their neighbors are all considered. The results indicate that 6-fold octahedral Ge and tetrahedral Ge have a similar local gap. However, 4-fold Ge at an octahedral site (4 neighbors with 90 degree angles) have both a shifted valence-band tail and conduction-band tail which may result in a bigger gap. Thus, from our result, $sp^3$ hybrids introduced by a Ge umbrella-flip may not be the reason for an increased gap in the amorphous phase, but the octahedral Ge existing in the amorphous phase at least would not increase the electronic gap. Analysis of Ge$_{1}$Sb$_{2}$Te$_{5}$ showed a similar result \citep{Raty09SSS}.

\begin{figure}[htbp]
\centering
\resizebox{0.9\columnwidth}{!}{
\includegraphics{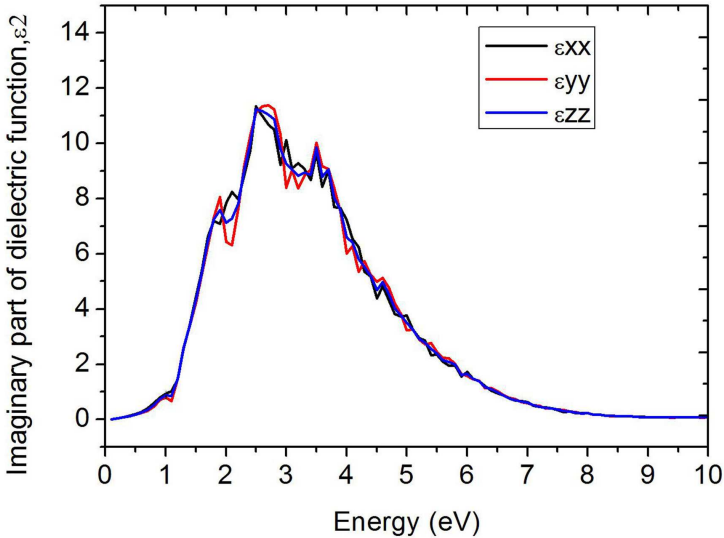}}
\caption{\label{figDiele} AC Dielectric function of a-\gst. Due to finite size effect, the calculation can not predict valid value for small $\omega$ ($\omega \lesssim 2 eV$).}
\end{figure}

Finally for this section, we show one last static property -- the dielectric function $\epsilon(\omega)$ in Fig.~\ref{figDiele}. The imaginary part of the dielectric function in three directions are plotted in Fig.~\ref{figDiele}. $\epsilon$ reaches its peak for an energy of about 2.5eV and this spectrum is comparable with both experimental and simulation result \citep{Wutting07nm,Raty09SSS}. Notice that due to the finite-size effect, the result is not valid for $\omega \rightarrow 0$. To obtain accurate results for small $\omega$, an extrapolation procedure is required.

\subsubsection{Dynamic Analysis}

\begin{figure}[htbp]
%\begin{center}
\centering
\resizebox{1.0\columnwidth}{!}{
\includegraphics{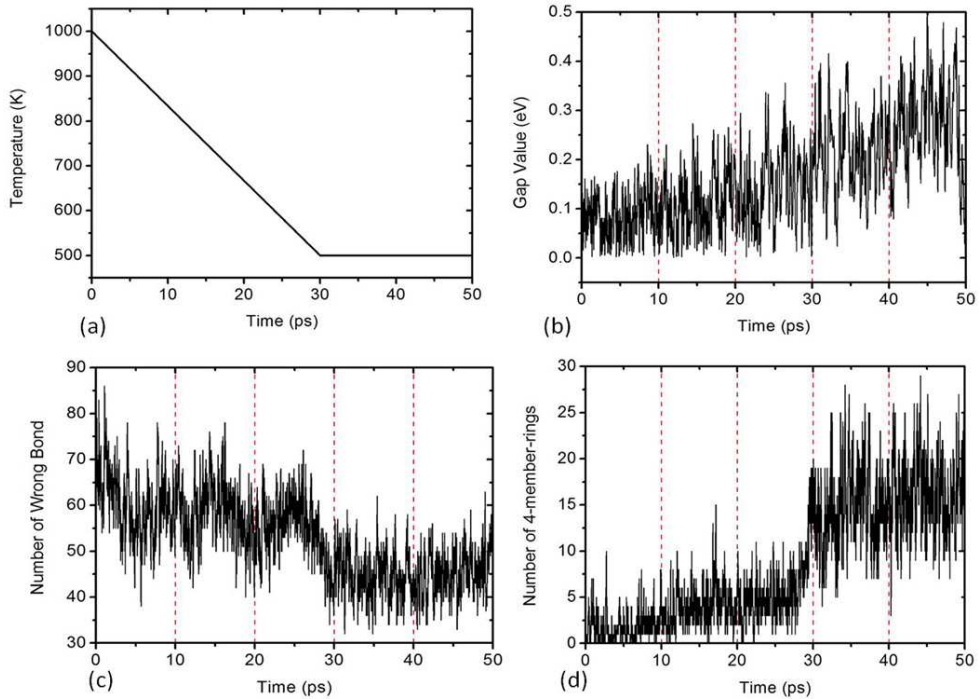}}
\caption{\label{figDynamic} Dynamic change of temperature, gap value, number of wrong bonds and squares (seeds).}
%\end{center}
\end{figure}

\begin{figure}[htbp]
%\begin{center}
\centering
\resizebox{0.9\columnwidth}{!}{
\includegraphics{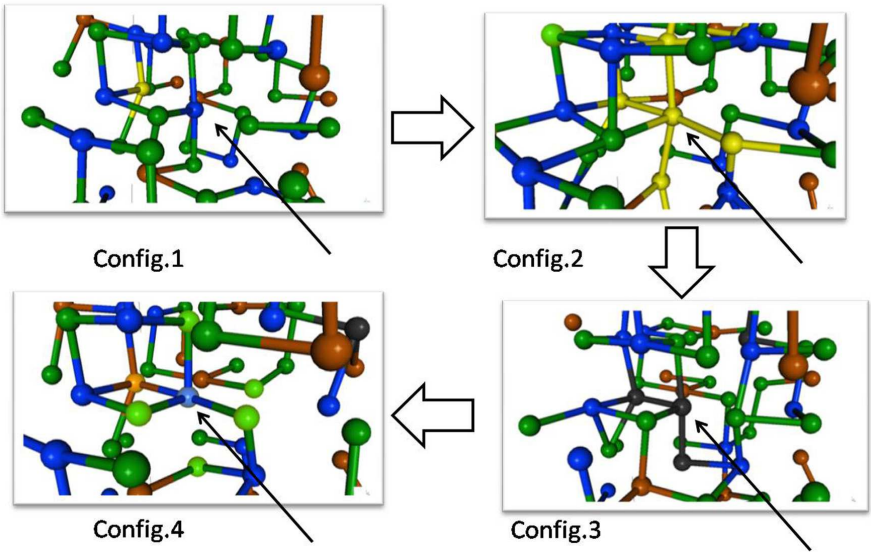}}
\caption{\label{figtranTOP} Snapshots of topology changes for one Ge atom and its six neighbors (Ge-blue, Sb-brown, Te-green). The central Ge atom is identified by black arrows. The valence-band tail states appear in Config.2 and are localized on yellow atoms. The conduction-band tail states appear in Config.3 and are localized on black atoms.}
%\end{center}
\end{figure}

Next, we performed a dynamic analysis for a-\gst. We tracked the structure and the electronic gap during a quench from 1000K to 500K with thermal equilibration at 500K (Fig.~\ref{figDynamic}). Significant structural changes started to occur after 24ps (the temperature was then near 640K). The number of homopolar bonds dropped, the number of 4-membered rings increased, and the mean coordination increased. The changes in topology are similar to those reported by \cite{Hed08} and all these shifts signal an increase of both chemical order and structural order. The electronic gap, which we take to be the difference between LUMO and HOMO levels, increased overall, but we observed that there are considerable fluctuations, even for the well-equilibrated system. Local geometry may have huge consequences on the gap.

\begin{figure}[htpb]
%\begin{center}
\centering
\resizebox{0.9\columnwidth}{!}{
\includegraphics{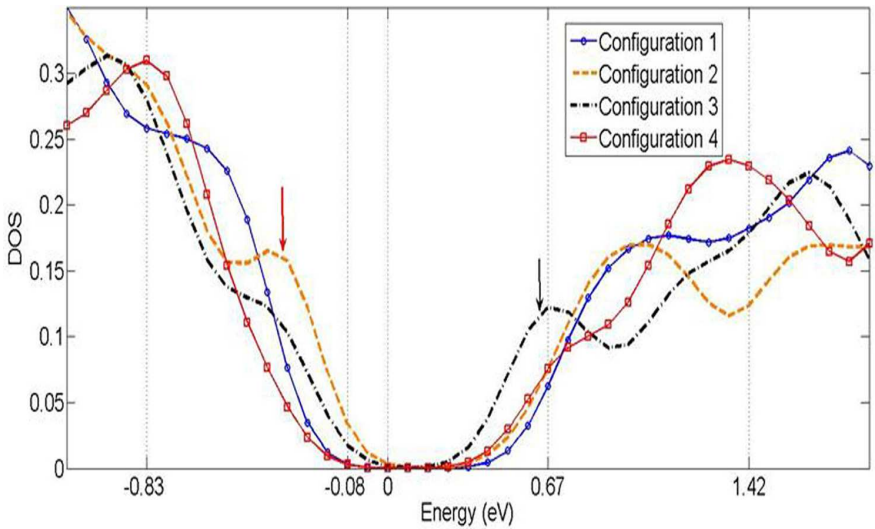}}
\caption{\label{figtranDOS} Instantaneous snapshots of EDOS correlated with the configurations of Fig.~\ref{figtranTOP}. A smaller gap appear in Config.2\&3. The valence-band tail states (orange arrow) are associated with yellow atoms in Config.2 of Fig.~\ref{figtranTOP}. The conduction-band tail states (black arrow) are associated with black atoms in Config.3 of Fig.~\ref{figtranTOP}. The Fermi level is at 0 eV.}
%\end{center}
\end{figure}
To study how changes in the local environment at a specific site affected the electronic gap, we tracked a specific unit in the system during equilibration and we show such an evolution for both the topology and the electronic structure in Fig.~\ref{figtranTOP} and Fig.~\ref{figtranDOS}. We mainly focused on one Ge atom which occupied a near-octahedral site (6 nearest Te neighbors with around 90 degree bond angles, indicated by a black arrow in Fig.~\ref{figtranTOP}) and its six nearest neighbors. We correlated their local bondings and electronic density of states for many time steps. Configurations 1 and 4 exhibit the biggest gap. However, at intermediate steps between configurations 1 to 4, tail states appear. At configuration 2, a valence-band tail state was present and it was mainly localized on the central Ge atom and four of its nearest neighbors (yellow atoms in Config.2 of Fig.~\ref{figtranTOP}); at configuration 3, a conduction-band tail appears, mainly localized on the center Ge atom and two of its nearest neighbors (black atoms in Config.3 of Fig.~\ref{figtranTOP}). We should emphasize that from configurations 1 to 4, the whole network did not experience a major change, but the electronic gap fluctuates. Thus, the appearance of valence-band and conduction-band tails are strongly associated with distortions at this Ge site. Our simulations emphasize the dynamic nature of the electronic band tails in \gst.

\subsubsection{Relaxation Analysis for Crystal Phase of \gst}

\begin{figure}[htbp]
%\begin{center}
\centering
\resizebox{0.9\columnwidth}{!}{
\includegraphics{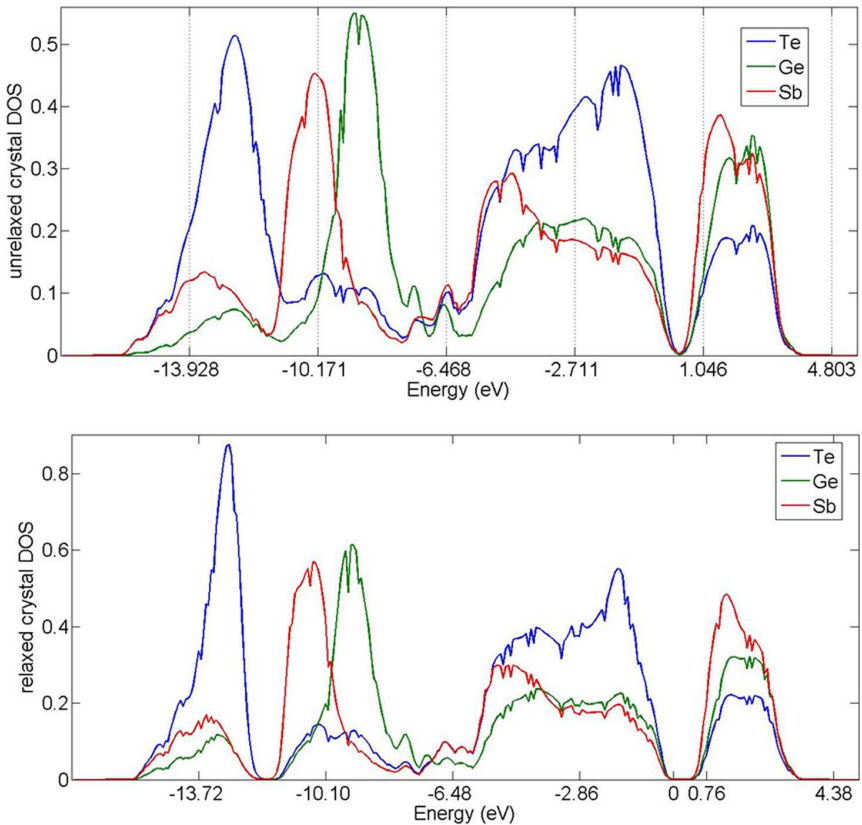}}
\caption{\label{figxtdos} Electronic density of states of crystal models projected onto different species of atoms. (Top-panel) Unrelaxed crystal model with vacancies. (Bottom-panel) Relaxed crystal model.}
%\end{center}
\end{figure}

\begin{figure}[htbp]
%\begin{center}
\centering
\resizebox{0.9\columnwidth}{!}{
\includegraphics{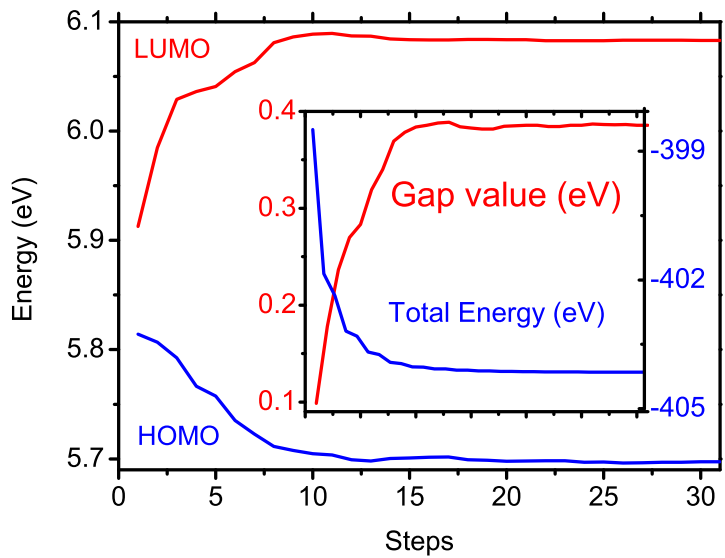}}
\caption{\label{figxtgap} Change of LUMO, HOMO level, gap value and total energy during relaxation.}
%\end{center}
\end{figure}

In this section, we discuss relaxation effects for crystalline \gst~ with 10\% vacancies. As mentioned above, the 108-atom cell was obtained based on the NaCl rock-salt structure. We show the electronic density of states for both unrelaxed and relaxed models in Fig.~\ref{figxtdos} obtained through HF calculations. For both models, Te atoms have a major effect on the valence tail which may be due to the vacancies; Sb atoms contribute more to conduction tail. We could see clearly that the electronic gap opened up after relaxation. Moreover, we tracked the dynamic change of the Highest Occupied Molecular Orbital (HOMO) and Lowest Unoccupied Molecular Orbital (LUMO) and plot them in Fig. ~\ref{figxtgap}. It is clear from the plot that the total energy is reduced and  both HOMO and LUMO levels are shifted. The HOMO level is pushed away by 0.1eV and the LUMO level is pushed up by 0.2eV.

Since the crystal model has 10\% vacancies, the relaxation actually introduced slight distortion into the network. The structural statistics indicate that the mean coordination of Te, Ge and Sb atoms all decreased. The mean coordination of Te are decreased from 4.8 to 4.28, Sb and Ge dropped from 6 to 5.47 and 5.23 correspondingly. The angle distribution, especially the X-Ge-X and X-Sb-X angle distributions, are also changed. This result indicates that the existence of vacancies and the distortion happened to the network will have a impact on gap. Thus, by controlling the concentration of vacancies and distortion, we may obtained different electronic gap values. This result is similar to results on other Ge-Sb-Te alloys \citep{Wutting07nm}.

\subsubsection{Conclusions on \gst}

We made \gst ~models with a `quench from melt' method. HF calculations give a 0.4eV electronic gap for the amorphous phase. We found that Te-p, Sb-p, Ge-p, Ge-s and Sb-s orbitals are most important to tail states. 6-fold octahedral Ge and 4-fold tetrahedral Ge give rise to similar gaps but 4-fold octahedral Ge results in a bigger gap with both shifted valence-band and conduction-band tails. The study also reveals a large fluctuation in gap value during thermal equilibration which is partially due to the appearance and disappearance of conduction-band and valence-band tail states. Such fluctuations could be associated with the local structural change/distortion of Ge atoms, which introduce localized tail states and have an impact on the electronic gap. Also, the relaxation analysis on crystal phase of \gst ~indicates that vacancies and distortions may play an important role in determining the electronic gap.

%% file: electrolyte.tex
\section{Electrolyte Materials} \label{gese}

\subsection{Background}

Electrolytes are materials with high ionic conductivity and high electrical resistivity. When doped with metals like Ag, chalcogenide glasses (e.g. Ge-Se) become solid electrolytes offering high ionic conductivities. Such electrolytes are getting attention for their technological importance with the application in "conducting bridge" (flash) memory devices \citep{mitkova}. It has been believed that a variety of different coordination patterns of Ag$^{+}$ ions in the glassy host with tiny energy differences is the basic reason why Ag$^{+}$ is mobile. Since the properties of chalcogenide glasses accrue from their structure, the knowledge of the structure of these glasses is an essential precursor for further study. From a material point of view it is interesting that an amorphous material should allow rapid motion of a transition metal ion through the network, and a great deal of energy has been devoted to understanding this phenomenon. Such diffusive processes in glasses have been studied for decades with a variety of experimental methods. There have also been several approaches to modeling such diffusive behavior.

There have been a wide range of experimental studies on the atomic structure of the amorphous state of electrolyte material and some computer simulations, typically on Ge-Se glasses doped with transition metals. Ge-Se-Ag based electrolyte materials have been studied experimentally using various techniques. For example, X-ray \citep{piarris} and neutron \citep{dejus,cuello} diffraction, and other experimental methods have been used to study the structure of Ge-Se-Ag glass. There have also been some computational studies to model the structure. Tafen et al. \citep{Tafen} reported two \textit{ab-initio} models; \Agi and \Agiii with short range order consistent with the experimental results. It has also been reported that Ag atoms prefer to sit at trapping center (TC) which is near the midpoint of a line joining two host atoms (Ge or Se) separated by a distance between 4.7 and 5.2 {\AA} with the bond length of Ag to the host atoms ranging between 2.4-2.6 {\AA} \citep{inam} for low Ag concentration. The simulation work has been also extended by introducing Cu into the network \citep{prasai}.

Beside structural studies, there have been quite a few studies on the conductivity of Ag doped chalcogenide glasses including both experimental and simulation work. Ag$_{x}$(GeSe$_{3}$)$_{100-x}$ glasses have been particularly studied for the ionic conductivity within a wide range of $x$ ($10$ to $25\%$). Ure{\~n}a et al. \citep{urena} predicted that the ionic conductivity follows an Arrhenius law. Tafen et al. presented a molecular dynamics(MD) simulation on Ag$_{x}$(GeSe$_{3}$)$_{100-x}$ with $x=10$ and $15\%$ \citep{Tafen} at different temperatures. In recent work, we have also presented a MD simulations on these glasses with the addition of Cu and illustrated the motion of the ions on the accessible time scales (tens of picoseconds) \citep{prasai}. Some of the results will be discussed in the following sections.

\subsection{Simulation of properties of Electrolyte Materials}

The models of Ag- and Cu-doped chalcogenide glasses discussed here were generated using the melt-quenching method. A cubic supercell is constructed with a fixed volume and a fixed number of atoms in order to reproduce the experimental density according to the desired stoichiometry. The atoms were randomly placed in the supercell with minimum acceptable distance between two atoms set to 2{\AA}. The calculations were carried out under periodic boundary condition using the Vienna \textit{Ab-initio} Simulation Package(VASP)\citep{vasp}, with Vanderbilt ultrasoft pseudopotentials. We used the local density approximation (LDA) for the exchange correlation energy. The details of the model generation can be found in the reference \cite{prasai}. Beside the models discussed there, two more models \Cuiii ~and \AgCu ~have been added to the discussion.

\subsection{Results and Discussion}

\subsubsection{Structural properties}

\begin{figure}
\centering
\resizebox{1.0\columnwidth}{!}{
\includegraphics{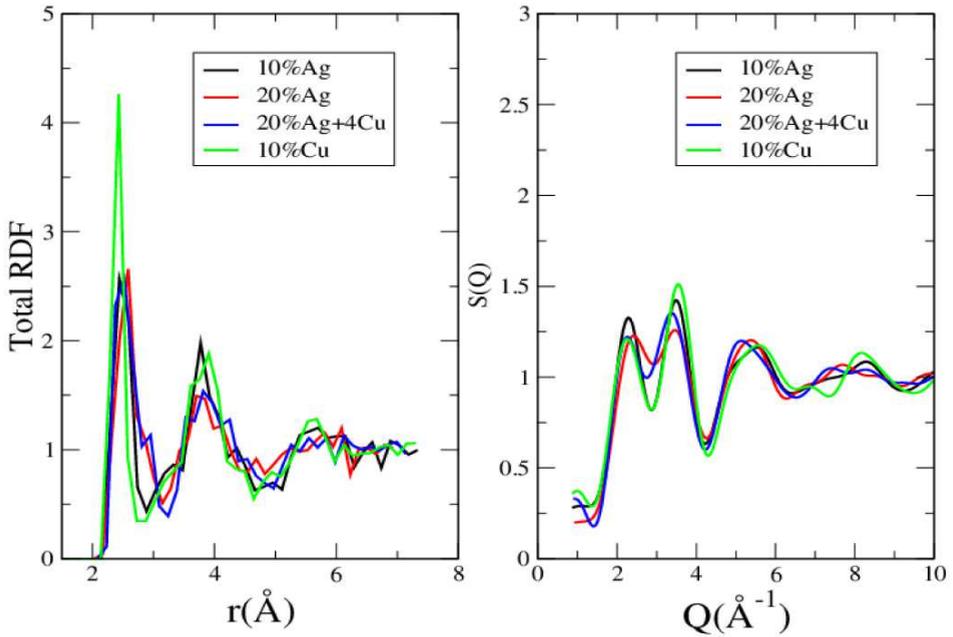}}
\caption{Comparison of total radial distribution functions and static structure factors for all amorphous models\label{figrdf}}
\end{figure}

\begin{figure}
\centering
\resizebox{1.0\columnwidth}{!}{
\includegraphics{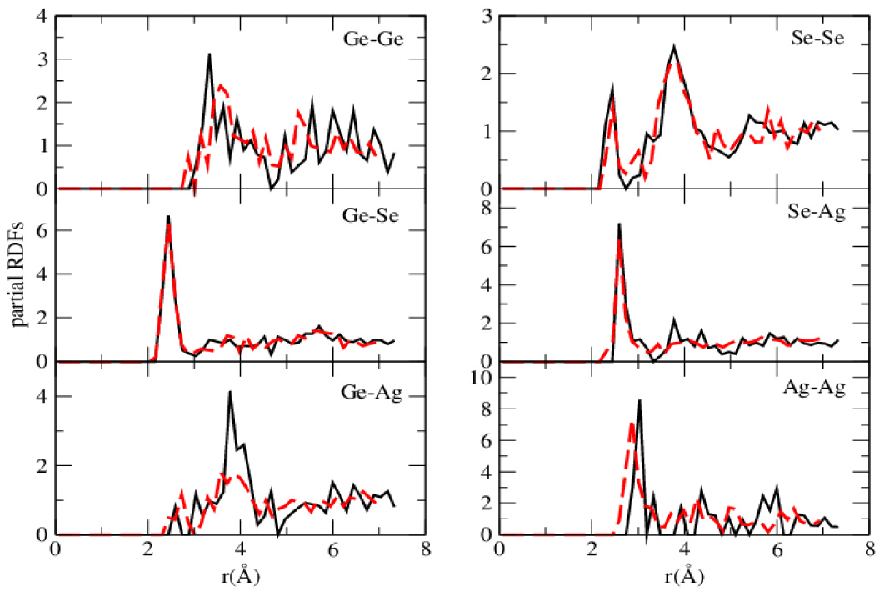}}
\caption{Partial radial distribution functions for amorphous \Agi(black) and \Agii(red/dashed line)
\label{figprdf}}
\end{figure}

\begin{figure}
\centering
\resizebox{1.0\columnwidth}{!}{
\includegraphics{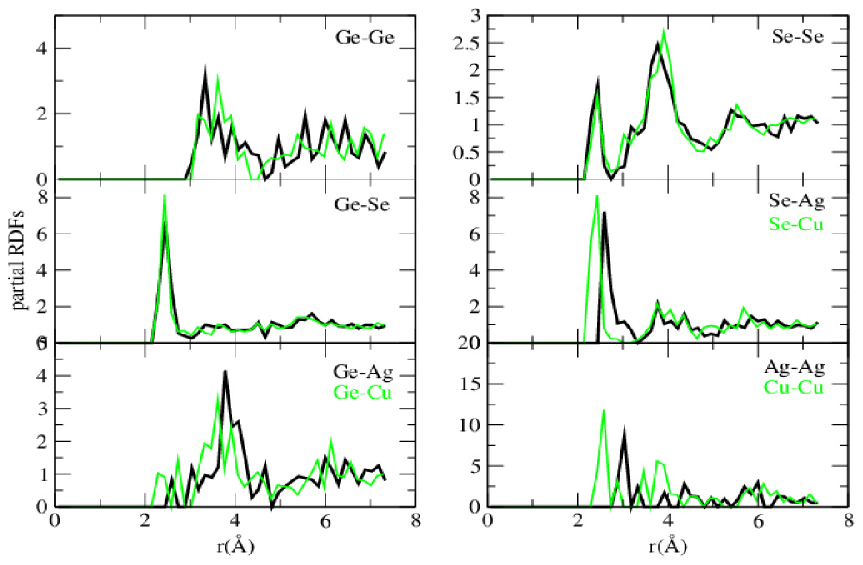}}
\caption{Partial radial distribution functions for amorphous \Agi(black) and \Cui(green/thin line)
\label{figprdf2}}
\end{figure}

Fig.~\ref{figrdf} shows the calculated total radial distribution functions (RDFs) and structure factors for the models; g-\Agi, g-\Agii, g-\Cui and g-\Cuii. The first peak of the RDF is the contribution from Ge-Se and Se-Se correlations whereas the second peak is due to Se-Se and Ge-Ag/Cu correlations(Fig.~\ref{figprdf} and Fig.~\ref{figprdf2}). There is not much variation in the short range order (SRO) i.e. nearest neighbor distance and second nearest neighbor distance for the different models. We observed a slight change in the nearest neighbor distance for the Ag rich model and Cu rich model. The average bond length and the mean coordination numbers are presented in Table \ref{tablecn}. We did not detect Ge-Ge bonds in any of our models as seen previously in g-\Agi \citep{Tafen}. We also observed that both Ag and Cu preferred to have Se as neighbor with only 16$\%$ of Cu/Ag bonded with Ge in our models. These results are very close to bond lengths measured by Piarristeguy et al. \citep{piarris}.  We also obtained the silver and copper coordination number for each model. The coordination number 3.1 of silver at 20$\%$ is as predicted(3.0) by Mitkova et al. \citep{mitkova2}. The coordination number 4.67 of copper at $10\%$ is much higher than 2.16 of silver (found to be 2.0 by Tafen et al. \citep{Tafen}) for the same concentration. We detected a few 3-fold Ge and 3 and 4 fold Se that we interpret as a structural defect in our models. Detailed bond parameters can be found in  \cite{prasai}.

We also compared the static structure factors for our models (Fig.~\ref{figrdf}). There is no significant change in the position of the first two peaks. We observed a weak peak in S(Q) slightly above 1 {\AA} $^{-1}$. This peak, which is a precursor to the first sharp diffraction peak (FSDP), varies as a function of Ag concentration and the peak disappears as Ag concentration increases, also shown by Piarristeguy et al. \citep{pairetal}. We did not observe any particular correlation contributing to this peak as the partial structure factors shows that the peak has contribution from all of the partials.  We compared partial structure factors for \Agi ~and \Cui ~and observed the only differences in correlation of Ag-Ag and Cu-Cu as well as in Se-Ag/Cu.

\begin{table}[tbp]
\begin{center}
\caption {Short range order; nearest neighbor distance(NN), next nearest neighbor distance(NNN) and mean coordination number(CN).}
\label{tablecn}
%\begin{ruledtabular}
\begin{tabular}{ccccccccc}
&NN({\AA})&&NNN({\AA})&&CN\\
\hline
\Agi&2.49&&3.75&&2.50\\
\Agii&2.51&&3.80&&2.92\\
\Cuii&2.45&&3.80&&2.9\\
\Cui&2.40&&3.83&&2.8\\
\end{tabular}
%\end{ruledtabular}
\end{center}
\end{table}

\begin{figure}
\centering
\resizebox{1.0\columnwidth}{!}{
\includegraphics{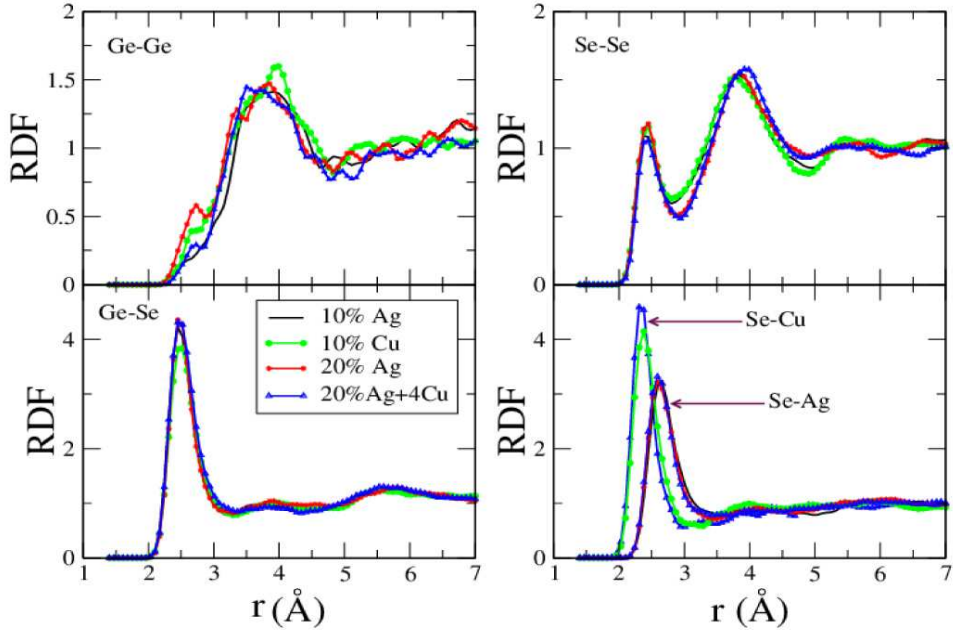}}
\caption{\label{figprdfliq}Comparison of partial radial distribution functions for all liquid models at 1000K}
\end{figure}

We performed thermal MD simulation at 1000K for 25ps in order to obtain well-equilibrated liquid systems. We calculated the total and partial radial distribution functions (RDF). The RDFs are averaged over the last 2.5 ps. The major peak positions in total RDF are 2.45 {\AA} for \Cui , 2.48 {\AA} for \Agi ~and 2.53 {\AA} for \Agii ~and \Cuii. We present partial radial distribution functions in Fig.~\ref{figprdfliq} showing Ge-Ge, Ge-Se, Se-Se and Se-Ag/Cu correlations. All of our models except \Agi (2.6{\AA}) confirm the presence of Ge-Ge homopolar bonds with peak position at 2.71 {\AA} in contrast with the glass. We also observed Se-Se and Ge-Se bond distances of 2.47{\AA} and  2.50 {\AA}, respectively. We observe no concentration dependence on the first peak position of Ge-Se,Se-Se and Se-Ag/Cu correlations. The major contribution to the first peak of the total RDF is from Ge-Se,Se-Se and Se-Ag/Cu correlations with Se-Ag/Cu correlation causing the shifts on the first peak positions. The second peak of the total RDF is mainly due to Se-Se correlation.

\subsubsection{Ion dynamics}
We studied the dynamics of Ag and Cu ions in the GeSe$_{3}$ host by computing the mean square displacement (MSD) for each atomic constituent as:
\begin{equation}
\langle r^{2}(t)\rangle_a = \frac{1}{N_{a}}\sum^{N_{a}}_{i=1}\langle\vert\vec{r_{i}}(t)-\vec{r_{i}}(0)\vert^{2}\rangle
\end{equation}
where the quantity in $\langle \rangle $ is the calculated statistical average over the particular atomic species $\alpha$.  We carried out constant temperature MD calculations at three different temperatures 300K, 700K and 1000K in order to study ion dynamics in our the amorphous as well as the liquid systems.

\paragraph{Amorphous Ge-Se-Cu-Ag}

\begin{figure}
\centering
\resizebox{1.0\columnwidth}{!}{
\includegraphics{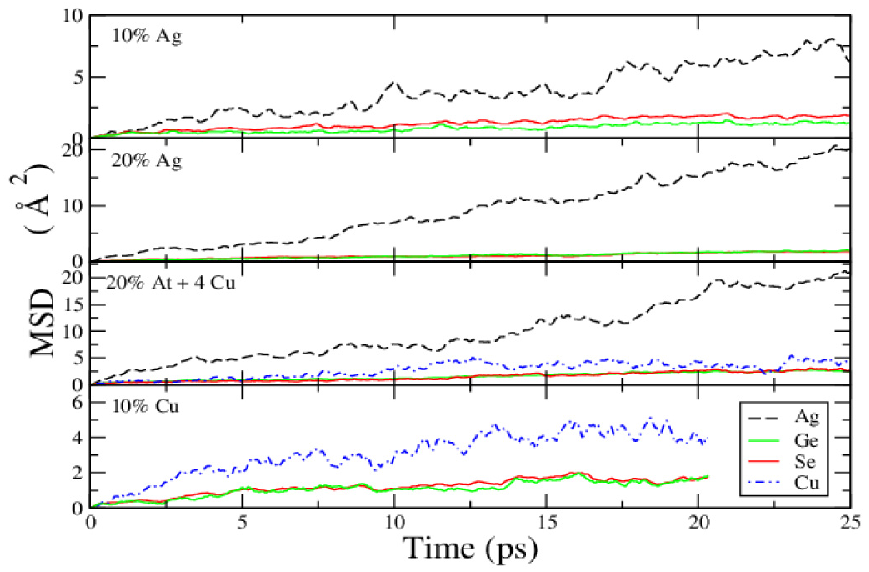}}
\caption{\label{figmsd}Mean square displacement of atoms in amorphous \Agi, \Agii, \Cuii and \Cui (top to bottom respectively) glasses at $T=700K$. Ag(black) Ge(green), Se(red) and Cu(blue)}
\end{figure}

\begin{figure}
\centering
\resizebox{1.0\columnwidth}{!}{
\includegraphics{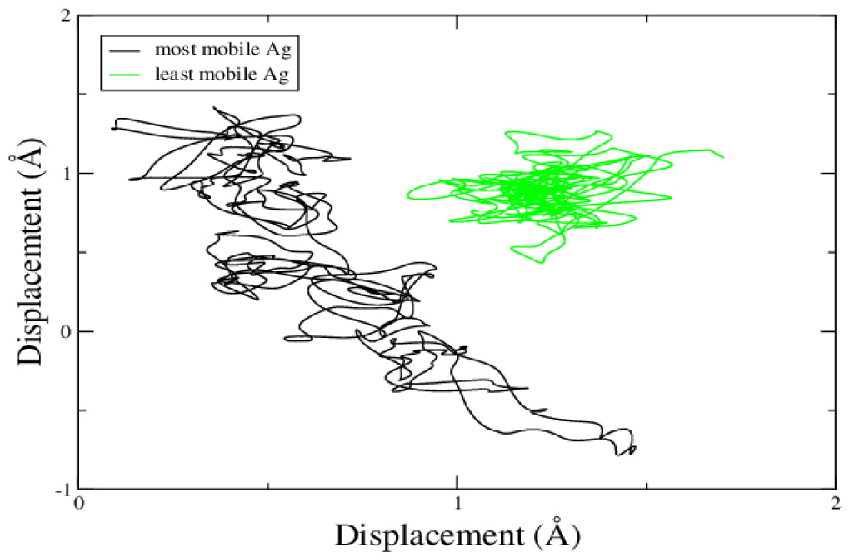}}
\caption{\label{figtrajag}Trajectories of the most and the least diffusive Ag ions at 700K as a function of time in amorphous \Agi.}
\end{figure}

\begin{figure}
\centering
\resizebox{1.0\columnwidth}{!}{
\includegraphics{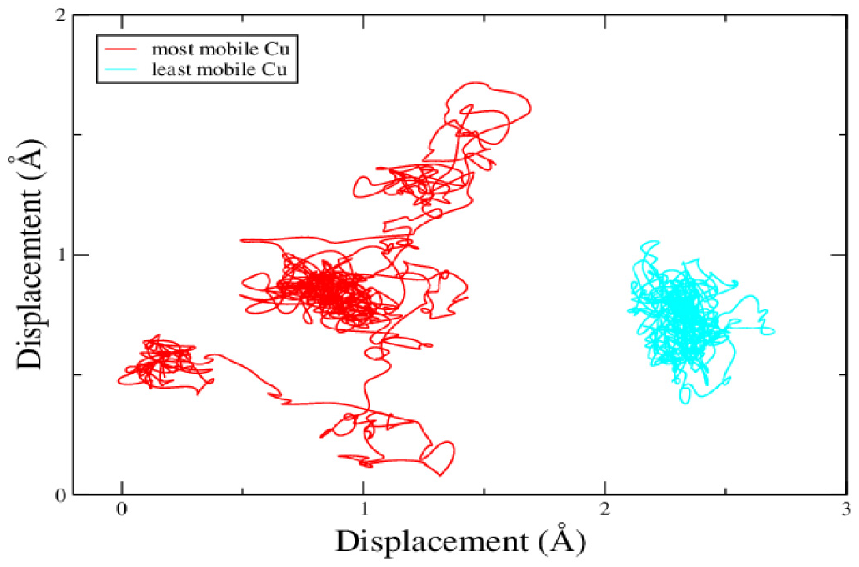}}
\caption{\label{figtrajcu}Trajectories of the most and the least diffusive Cu ions at 700K as a function of time in amorphous \Cui.}
\end{figure}

As expected, at 300K none of the ions showed measurable diffusion. In order to investigate the diffusion in the solid place, we chose $T=700K$ and present the MSD for each species for each system calculated at this temperature in Fig.~\ref{figmsd}. At 700K Ag$^{+}$ ions show significant diffusion consistent with the previous result\citep{Tafen} in contrast to Cu ions that do not diffuse much. To elucidate the diffusion of these ions we examine the trajectories for 20ps. Fig.~\ref{figtrajag} and ~\ref{figtrajcu} show two dimensional projections of the trajectories of the most and the least diffusive ions in \Agi ~and \Cui. The trajectories illustrate the wide range of diffusion for the ions with displacement ranging 1{\AA}-3.87{\AA} in \Agi, 2{\AA}-6.71{\AA} in \Agii  ~and 1{\AA}-3.74{\AA} in \Cui. For the mixed-ion model \Cuii, this displacement ranges between 1.73{\AA}-2.82{\AA} for Cu and 1.41{\AA} - 8.06{\AA} for Ag. For Ag rich models more than 60$\%$ of the ions exhibit displacements greater than the average displacement (2.36{\AA} in \Agi ~and 4.47{\AA} in \Agii) whereas for Cu, the majority has displacement smaller than the average(2.11{\AA}).  The wide range of diffusion can be attributed to variation in the local environment of the ions. To illustrate this we calculated the local densities of the most and the least mobile ions. We employed a sphere of radius 5.0{\AA} around the ion and calculated the mean density of atoms inside the sphere.  We observed that the most diffusive ion is located in the region with lower local density. In other words the most mobile ions have the wider variation of the local density as compared to that of the least mobile ion.

\paragraph{Liquid Ge-Se-Cu-Ag}

\begin{figure}
\centering
\resizebox{1.0\columnwidth}{!}{
\includegraphics{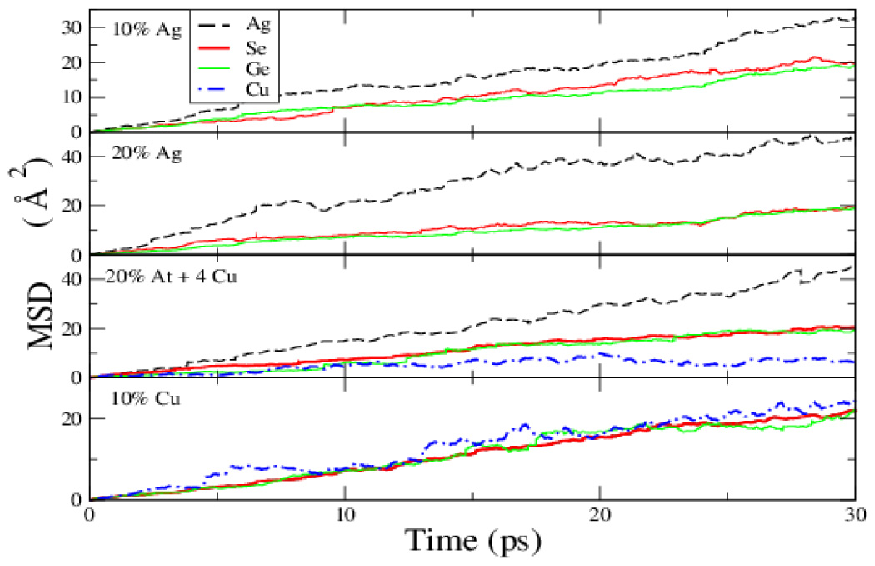}}
\caption{\label{figmsdliq}Mean square displacement of atoms in liquid \Agi, \Agii, \Cuii and \Cui (top to bottom respectively) glasses at $T=1000K$. Ag(black) Ge(green), Se(red) and Cu(blue)}
\end{figure}

One of the essential properties of a liquid is the high diffusivity of atoms in the system. To illustrate this, we calculated the mean square displacements for each species at 1000K in all of our models. The diffusion plots as presented in Fig.~\ref{figmsdliq}  shows that the MSD of each species increases rapidly as compared to that at 700K. We observe Ag diffusion still significantly larger than the host particles however; Ge and Se atoms are also diffusing rapidly. As before Cu still does not show high diffusion as Ag does compared to the host atoms.

\begin{figure}
\centering
\resizebox{1.0\columnwidth}{!}{
\includegraphics{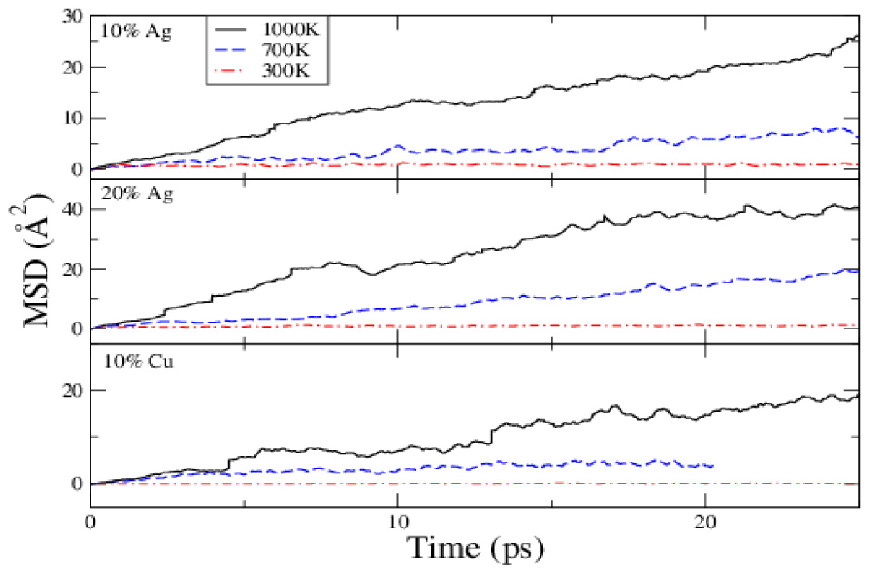}}
\caption{\label{figmsdcomp} Temperature dependence of conductivity of ions for different models)}
\end{figure}

Based on the plots we calculated diffusion coefficients using Einstein relation \citep{einstein}. The Einstein relation for self-diffusion is given by:
\begin{equation}
\langle\vert\vec{r_{i}}(t)-\vec{r_{i}}(0)\vert^{2}\rangle=6Dt + C
\end{equation}
where $C$ is a constant and $D$ is the self-diffusion coefficient. The conductivity can be calculated from the equation
\begin{equation}
\sigma=\frac{ne^{2}D}{k_{B}T}
\end{equation}
where $n$ is the number density of ions.
The temperature dependence of the diffusion is shown in Fig.~\ref{figmsdcomp} and the values of diffusion coefficients and conductivities at different temperatures are presented in Table ~\ref{tablecond}. We did not find experimental results for the conductivity of Cu ions; however Ag conductivity is close to ones reported by Ure{\~n}a et al.\citep{urena}.

\begin{table}[tbp]
\begin{center}
\caption {Self diffusion coefficient D and conductivity $\sigma$  at 300K, 700K and 1000K for \Agi (10$\%$Ag), \Agii (20$\%$Ag), \Cui (10$\%$Cu) and \Cuii (0.77$\%$Cu)}
\label{tablecond}%
\begin{tabular}{ccccccccc}
&T(K)&D(cm$^{2}$/s)&&$\sigma$(Scm$^{-1}$)\\
&&&&This work&Expt.\cite{urena}\\
\hline
10$\%$Ag&300K&1.15$\times$10$^{-9}$&&2.63$\times$10$^{-5}$&1.3$\times$10$^{-5}$\\
&700K&4.53$\times$10$^{-6}$&&4.44$\times$10$^{-2}$&2.07$\times$10$^{-2}$\\
&1000K&1.23$\times$10$^{-5}$&&8.45$\times$10$^{-2}$&8.98$\times$10$^{-2}$\\
20$\%$&300K&1.16$\times$10$^{-8}$&&5.3$\times$10$^{-4}$&7.5$\times$10$^{-5}$\\
&700K&1.20$\times$10$^{-5}$&&2.35$\times$10$^{-1}$&6.57$\times$10$^{-2}$\\
&1000K&2.53$\times$10$^{-5}$&&3.47$\times$10$^{-1}$&2.584$\times$10$^{-1}$\\
10$\%$Cu&300K&7.3$\times$10$^{-10}$&&1.67$\times$10$^{-5}$\\
&700K&3.3$\times$10$^{-6}$&&3.23$\times$10$^{-2}$\\
&1000K&1.13$\times$10$^{-5}$&&7.75$\times$10$^{-2}$\\
0.77$\%$Cu&300K&D$_{Ag}$=1.06$\times$10$^{-8}$&&4.85$\times$10$^{-4}$\\
&&D$_{Cu}$=7.16$\times$10$^{-9}$&&1.63$\times$10$^{-5}$\\
&700K&D$_{Ag}$=1.30$\times$10$^{-5}$&&2.54$\times$10$^{-1}$\\
&&D$_{Cu}$=1.16$\times$10$^{-6}$&&3.8$\times$10$^{-3}$\\
&1000K&D$_{Ag}$=2.42$\times$10$^{-5}$&&3.32$\times$10$^{-1}$\\
&&D$_{Cu}$=5.24$\times$10$^{-6}$&&1.2$\times$10$^{-2}$\\
\end{tabular}
\end{center}
\end{table}

\subsubsection{Trap centers and hopping of ions}

\begin{figure}
\centering
\resizebox{1.0\columnwidth}{!}{
\includegraphics{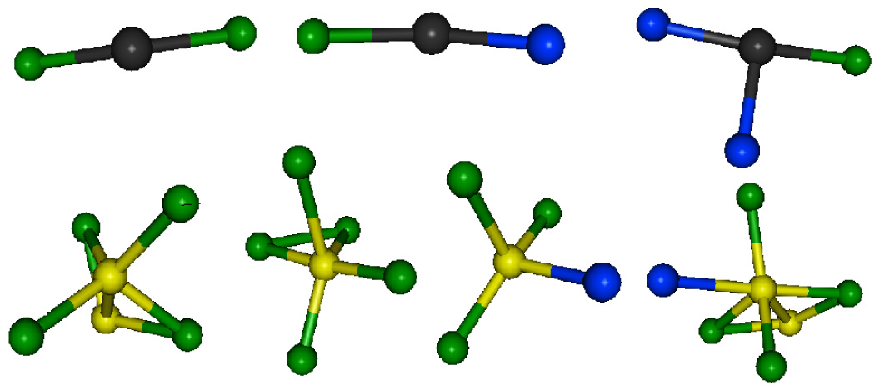}}
\caption{\label{figtc} Local environments of Ag atoms(top) and Cu atoms (bottom). Black, green,blue and yellow colored atoms respectively represent Ag, Se, Ge and Cu)}
\end{figure}

To illustrate the different ionic transport properties of Ag and Cu, it is essential to study the local environment of Ag and Cu in our models. Fig.~\ref{figtc} shows the local environment for Ag and Cu in  \Agi ~and \Cui respectively. In the relaxed networks, most of the Ag ions(58.3$\%$) are found to occupy the trap centers, between two of the host sites as also predicted by the previous workers \citep{Tafen,inam} but this is not the same case with Cu. Cu is always surrounded by more than two host atoms that makes the traps for Cu more rigid than for Ag. In Ag rich systems at 300K, we observed that Ag is basically trapped with only a few hopping events. At 700K the lifetime of the trap decreases and hopping occurs. We observed the lifetime of the traps varying from 1ps - 3.5ps. However at 1000K we failed to observe well defined hopping events because of the high the diffusion of the host itself. In the Cu rich system the story is completely different. Even at 700K we could observe only a few hopping events with much larger trap life time. It has also been shown by previous workers that the nature of trap or cage depends mainly on coordination number, nearest neighboring distance and angular distribution of the nearest neighbors \citep{kraemer}. The low coordination number of Ag makes it easy to escape the trap whereas for Cu, high coordination number, smaller neighbor distance and a more uniform angular distribution makes it more difficult to escape from the trap.

\subsubsection{Mixed Ion Conductivity}

\begin{figure}
\centering
\resizebox{0.9\columnwidth}{!}{
\includegraphics{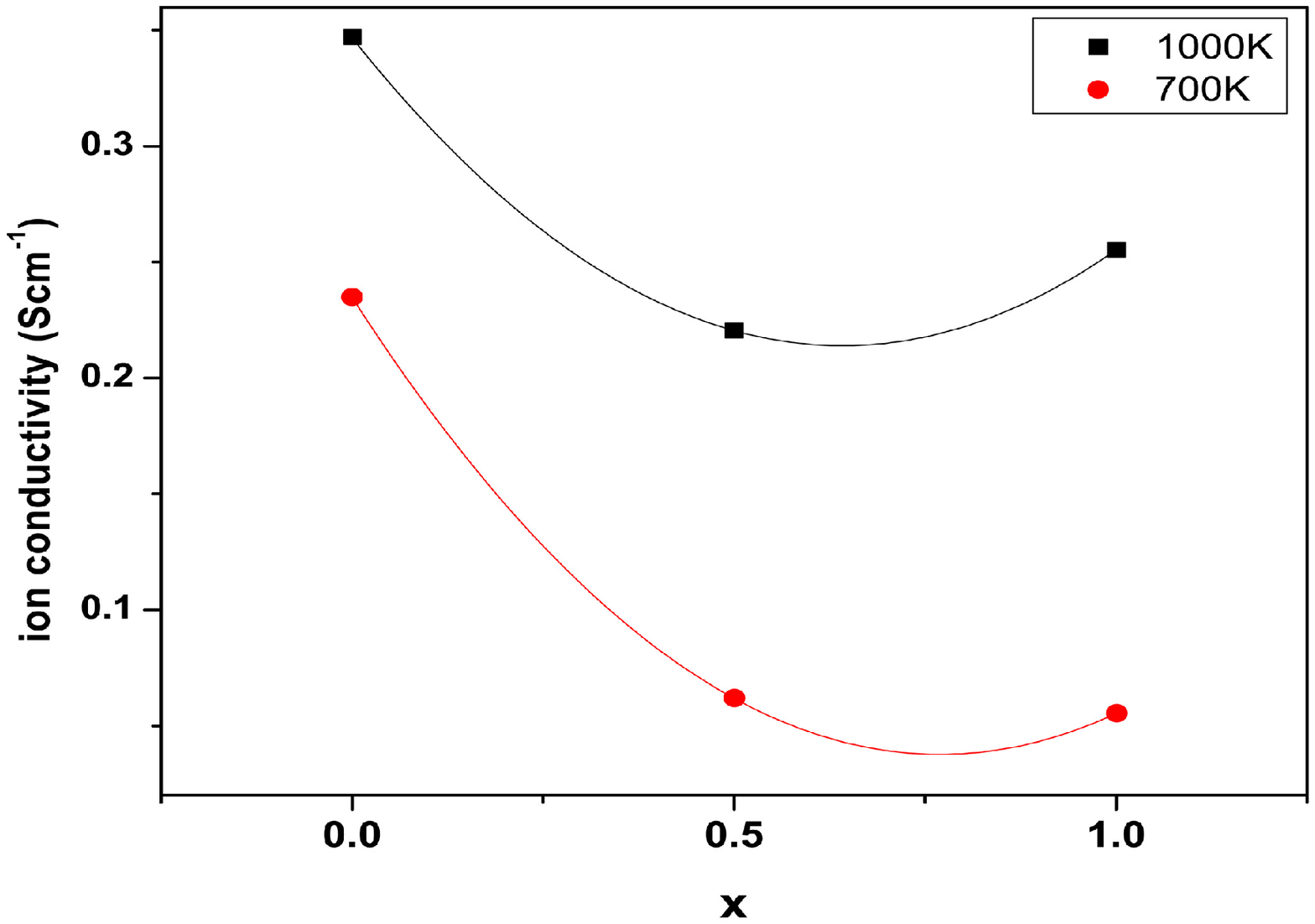}}
\caption{\label{figmixedion} Mixed ion effect: comparison of ion conductivities (GeSe$_{3}$)$_{0.8}$(Ag$_{1-x}$Cu$_{x}$)$_{0.2}$ glasses as a function of x.}
\end{figure}

One big challenge in these materials is to fully understand the effect on the dynamic properties such as ionic conductivity when one of the mobile ion is partially substituted by another type of mobile ion. There is a non-linear change in ionic mobility when two or more than two types of mobile ions are mixed in ion conducting glasses and crystals, and the effect is known as mixed ion effect. This section reveals that the mixed ion effect in Ag and Cu doped GeSe$_{3}$ glasses is present in our simulation. Constant temperature MD simulations were carried out in (GeSe$_{3}$)$_{0.8}$(Ag$_{1-x}$Cu$_{x}$)$_{0.2}$ where $x= 0, 0.5$ and $1$ at two different temperatures of 700K and 1000K. The calculated ion conductivities are presented in Fig.~\ref{figmixedion}. The figure shows a drastic drop in the ionic conductivity when both Ag and Cu ions are present in the system. This result implies a mixed ion effect in Ag/Cu doped chalcogenide glass, where Ag$^{+}$ conduction is greatly reduced by the presence of Cu$^{+}$. It is encouraging to see a mixed-ion effect in our simulations; its atomistic origin is under study.

\subsection{Conclusion: Fast ion conducting glasses}

We prepared different Ag and Cu doped GeSe$_{3}$ glass and liquid models by \textit{ab initio} simulation using the 'melt-quench' method and analyzed their structural and electronic properties. We also simulated dynamics of Ag and Cu ions using molecular dynamics. We were able to reproduce structural data as provided by X-ray diffraction. From the electronic density of state we observed that the increase in Ag concentration widens the optical gap whereas increase in Cu concentration narrows the gap. We were also able to see the metallic behavior for the liquid systems with the gap closing completely at 1000K. We were able to show the diffusion of the ions even in our time scale and predict the conductivity close to the experimental data. We also studied the trap and found that Cu traps are more rigid that those for Ag making very hard for Cu to diffuse.